\documentclass{emulateapj}
\pdfoutput=0
\usepackage{graphicx}
\usepackage{natbib}

\singlespace
\shortauthors{Kogut}
\shorttitle{Synchrotron Spectrum from 22 MHz to 23 GHz}
\slugcomment{Accepted by The Astrophysical Journal}

\begin{document}

% ---------------------Title ---------------------
\title{Synchrotron Spectral Curvature from 22 MHz to 23 GHz}

% --------------------- Author list ---------------------
\author{ A. Kogut\altaffilmark{1} }
\altaffiltext{1}{Code 665, Goddard Space Flight Center, Greenbelt, MD 20771}
\email{Alan.J.Kogut@nasa.gov}

% --------------------- Abstract ---------------------
\begin{abstract}
We combine surveys of the radio sky at frequencies
22 MHz to 1.4 GHz
with data from the ARCADE-2 instrument
at frequencies 3 to 10 GHz
to characterize the frequency spectrum
of diffuse synchrotron emission
in the Galaxy.
The radio spectrum
steepens with frequency
from 22 MHz to 10 GHz.
The projected spectral index at 23 GHz
derived from the low-frequency data
agrees well with independent measurements
using only data at frequencies 
23 GHz and above.
Comparing the spectral index at 23 GHz
to the value from previously published analyses
allows extension of the model to higher frequencies.
The combined data are consistent with a power-law index
$\beta = -2.64 \pm 0.03$ at 0.31 GHz,
steepening by an amount
$\Delta \beta = 0.07$ every octave in frequency.
Comparison of the radio data to models 
including the cosmic ray energy spectrum
suggests that any break in the synchrotron spectrum
must occur at frequencies above 23 GHz.

\end{abstract}

\keywords{
radio continuum: general,
radiation mechanisms: non-thermal
}

% --------------------- Main text ---------------------

% \clearpage
% \newpage

\section{Introduction}
Synchrotron emission from relativistic cosmic ray electrons
accelerated in the Galactic magnetic field
dominates the diffuse radio continuum
at frequencies below 1 GHz.
It is an important foreground contaminant
for measurements of the cosmic microwave background radiation,
and also serves to probe
the Galactic magnetic field
and cosmic ray distributions.
Measurements of the synchrotron frequency spectrum
are thus of interest to several areas in astrophysics.

An isotropic distribution of relativistic electrons
at a single energy
$E = \gamma m c^2$
propagating in a uniform magnetic field $B$
has emissivity
\begin{equation}
\epsilon(\nu) = 
	\frac{ \sqrt{3} e^3}{m c^2}
	B \sin\alpha
	F(x) ~,
\label{synch_emissivity_eq}
\end{equation}
where
$\alpha$ is the pitch angle between the magnetic field
and the line of sight, 
and
\begin{equation}
F(x) = x \int^\infty_x K_{5/3}(x^{\prime}) dx^{\prime}
\label{bessel_eq}
\end{equation}
is defined in terms of the 
modified Bessel function of order $5/3$
with variable
$x = \nu / \nu_c$
and
\begin{equation}
\nu_c = \frac{3}{4 \pi}
	\frac{e}{m c}
	\gamma^2 
	B \sin\alpha
\label{synch_crit_freq_def}
\end{equation}
\citep{schwinger:1949,
westfold:1959,
oster:1961}.
For a power-law distribution of electron energies
$N(E)~\propto~E^{p}$
propagating in a uniform magnetic field,
the synchrotron emission is also a power law,
\begin{equation}
T_A(\nu) \propto \nu^{\beta}
\label{synch_power_law}
\end{equation}
where
$T_A$ is antenna temperature,
$\nu$ is the radiation frequency,
and
\begin{equation}
\beta = \frac{p-3}{2}
\label{synch_index}
\end{equation}
\citep{rybicki/lightman:1979}.

Measurements of the synchrotron spectral index
provide important input for models of cosmic ray propagation.
Solar modulation reduces the local cosmic ray electron density
for electron energies below a few GeV
so that synchrotron emission provides
the most direct probe of low-energy cosmic rays.
Measurements of the cosmic ray spectrum above a few GeV
in turn 
inform models of the high-frequency synchrotron spectrum.
Energy losses from cosmic ray propagation 
steepen the cosmic ray spectrum,
increasing $p$ toward higher energies.
The observed steepening 
from
$p \sim -2.6$ at 5 GeV
to
$p \sim -3.2$ at 50 GeV
predicts a corresponding steepening in the synchrotron spectrum
from $\beta \sim -2.8$ at 1 GHz
to $\beta \sim -3.1$ at 100 GHz
\citep{strong/etal:2007}.

Comparison of the cosmic ray spectra
to the predicted synchrotron spectrum
is complicated by confusion from competing radio emission sources.
The diffuse radio continuum
is a superposition of
the cosmic microwave background,
synchrotron emission,
free-free emission from the warm ionized interstellar medium,
and emission from interstellar dust.
A number of authors have attempted to disentangle
the various emission sources
to determine the synchrotron spectral index
(for a recent review see Appendix A of
\citet{strong/etal:2011}).
Despite some discrepant results,
the general trend shows a steepening of the synchrotron spectrum
from $\beta \sim -2.5$ at 22 MHz
to $\beta \sim -3.0$ above 23 GHz,
in rough agreement with the observed cosmic ray spectra.

Several factors contribute to the observed scatter
in estimates of the synchrotron spectral index.
Most estimates,
particularly those below 23 GHz,
assume a power-law spectrum for synchrotron
and do not explicitly model spectral steepening.
Comparisons between closely-separated frequencies
more accurately reflect the local synchrotron spectrum,
but have larger uncertainties from competing emission sources
or measurement offsets.
Analyses with broader frequency coverage
reduce foreground and offset uncertainties
but average over any spectral steepening.

Two additional effects are important
for analyses including data from the
Wilkinson Microwave Anisotropy Probe (WMAP)
at frequencies 23 to 94 GHz.
A growing body of evidence suggests
that a substantial fraction of the diffuse continuum
near 23 GHz
consists of
electric dipole radiation from a population 
of small, rapidly spinning dust grains
\citep{kogut/etal:1996,
oliveira-costa/etal:1997,
oliveira-costa/etal:2004,
miville/etal:2008,
dobler/finkbeiner:2008,
ysard/etal:2010,
kogut/etal:2011,
gold/etal:2011,
planck/xx:2011}.
Spinning dust emission is expected to peak
at frequencies near 20 GHz
\citep{draine/lazarian:1998,
ali-haimoud/etal:2009,
hoang/draine/lazarian:2010,
ysard/verstraete:2010}.
Analyses that ignore this component
to attribute the observed emission only to synchrotron radiation
tend to over-predict the synchrotron amplitude 
at frequencies near 23 GHz,
biasing the derived spectral index
to flatter values when comparing to lower frequencies
and steeper values when comparing to higher frequencies.

A second systematic error can result from
improper treatment of offsets in the data.
Measurements from radio surveys
at frequencies below 20 GHz
generally include the absolute intensity
(zero level) of the sky.
The WMAP differential radiometers
are insensitive to 
any constant (monopole) intensity on the sky;
the zero level of the WMAP sky maps
is set so that the map intensity in the Galactic polar caps
matches a cosecant fit to the mid-latitude sky
\citep{bennett/etal:2003,
hinshaw/etal:2009}.
Analyses that directly compare low-frequency radio surveys
to the WMAP data
without subtracting a monopole component
from the radio data
will miss the corresponding emission in the WMAP bands,
biasing the derived spectral index to steeper values.

%---------------------------------------------------------
% Figure 1: Toy model showing bias induced from zero level effects
%---------------------------------------------------------
\begin{figure}[b]
\includegraphics[width=2.9in,angle=90]{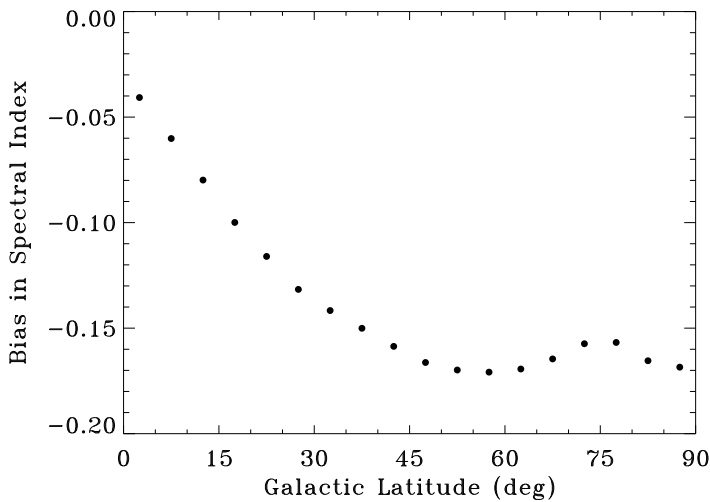}
\caption{Toy model showing effect 
of improper zero level subtraction on
the derived spectral index between 408 MHz and 23 GHz.
\label{toy_model_fig} 
}
\end{figure}
%---------------------------------------------------------

The zero level bias can be significant.
The \citet{haslam/etal:1981} survey
at 408 MHz is commonly used to model synchrotron emission.
The North Galactic pole
has measured temperature
$19 \pm 3$ K at 408 MHz,
while a $\csc|b|$ fit to the same 408 MHz map
predicts a polar contribution of only
$5.1 \pm 0.6$ K.
Similar results apply to the Southern hemisphere,
where the measured polar cap temperature of
$21 \pm 3$ K
significantly exceeds the value
$4.0 \pm 0.5$ K
obtained from a $\csc|b|$ fit.
Only 2.7 K of the difference
can be attributed
to emission from the cosmic microwave background,
leaving a large residual.

Figure \ref{toy_model_fig}
illustrates the bias induced
by including this residual at 408 MHz
but excluding it from the WMAP data.
We take the 408 MHz map,
remove the 2.7 K CMB monopole,
and scale the remaining (radio) emission
to 23 GHz using a power-law index $\beta = -2.7$
to produce a toy model of the radio sky at 23 GHz.
Following the WMAP processing, 
we then remove a monopole from the scaled map
so that the map temperature in the south polar cap
matches the $\csc|b|$ fit.
We then compute the bias in apparent spectral index
by comparing the 408 MHz map
to the scaled 23 GHz map
before and after removing the scaled monopole.
Figure \ref{toy_model_fig} shows the bias in spectral index at 23 GHz, 
binned by Galactic latitude.
Dis-similar treatment of the map zero level
creates a spatially varying bias $\Delta \beta \approx 0.15$,
comparable to the total spectral steepening
predicted by the measured cosmic ray spectra.
The bias is largest in regions
where the sky brightness is faintest,
at high latitudes or away from the Galactic center.

%---------------------------------------------------------
% Table 1: Sky Map Summary
%---------------------------------------------------------
\begin{table}[t]
\begin{center}
\caption{Sky Surveys Used for Synchrotron Analysis}
\label{sky_map_table}
\begin{tabular}{c c c c}
\tableline
\tableline
Frequency & Calibration & Offset	  & Relative \\
(GHz)     & Uncertainty & Uncertainty (K) & Uncertainty$^a$  \\
\tableline
0.022	  & 0.05  & 5000	& 0.15 \\
0.045	  & 0.10  & 250		& 0.11 \\
0.408	  & 0.10  & 3.0		& 0.17 \\
1.420	  & 0.05  & 0.5		& 0.63 \\
3.20	  & 0.001 & 0.011	& 0.10 \\
3.41	  & 0.001 & 0.006	& 0.07 \\
7.98	  & 0.001 & 0.036	& 0.89 \\
8.33	  & 0.001 & 0.042	& 2.64 \\
9.72	  & 0.001 & 0.003	& 0.34 \\
10.49	  & 0.001 & 0.002	& 0.27 \\
\tableline
\multicolumn{4}{c}{$^a$Quadrature sum of calibration and offset uncertainties,} \\
\multicolumn{4}{c}{divided by the mid-latitude sky temperature.} \\
\end{tabular}
\end{center}
\end{table}
%---------------------------------------------------------

Measurement uncertainties in the absolute level of the sky brightness
can also introduce bias in estimates of the synchrotron spectral index.
Many of the low-frequency surveys have uncertainty
in the measured zero level
approaching 30\% of the polar cap brightness.
As with the toy model above,
such measurement errors introduce spatially dependent biases
that are largest where the sky brightness is faintest.
Minimizing uncertainty in the derived synchrotron spectrum
requires a combination of
measurements with good sky coverage
and good control of offset uncertainty
at frequencies where competing emission sources are faint.
No such ideal data set yet exists.
In this paper,
we model the synchrotron spectral index and curvature
using low-frequency radio surveys
with high sky coverage but large offset uncertainty,
combined with
higher-frequency measurements
with limited sky coverage but still useful offset uncertainty.

\section{Sky Maps}

We model synchrotron emission
using radio surveys at 
22 MHz
\citep{roger/etal:1999},
45 MHz
\citep{maeda/etal:1999,alvarez/etal:1997},
408 MHz
\citep{haslam/etal:1981},
and 1420 MHz
\citep{reich/etal:2001,reich/reich:1986}.
These surveys have full or nearly-full sky coverage
at frequencies where 
Galactic radio emission is significant,
with gain and zero-level systematics controlled 
at the 10--20\% level.
We supplement the radio surveys
with sky maps from the
Absolute Radiometer for Cosmology, Astrophysics, and Diffuse Emission
(ARCADE 2) instrument\footnote{
The ARCADE data are available at
the Legacy Archive for Microwave Background Data Analysis,
{\tt http://lambda.gsfc.nasa.gov}}
at 3, 8, and 10 GHz
\citep{kogut/etal:2011}.
The ARCADE 2 data observe both the Galactic plane
and mid-latitude regions ($|b| < 40\arcdeg$)
with sufficient control of zero-level uncertainty 
to constrain the synchrotron curvature
relative to the lower-frequency radio surveys.

Table 1 summarizes the input sky maps.
The increase in the offset uncertainty 
at low frequency is compensated 
by a corresponding increase in sky brightness.
The final column shows the relative measurement uncertainty
for a mid-latitude region,
defined 
as the ratio of the combined offset and calibration uncertainty
to the measured brightness at $(l,b) = (17\arcdeg, -35\arcdeg)$
after removing the CMB monopole.
The selected maps provide
roughly uniform relative sensitivity to synchrotron emission
over 2.5 decades of frequency.

We convert all maps to units of antenna temperature
and subtract the CMB monopole
at (thermodynamic) temperature 2.725 K
from the measured sky temperatures.
We then convolve each map to the 11\fdg6 ~angular resolution
of the ARCADE 2 instrument.
At frequencies of 10 GHz and below,
both thermal dust emission
and spinning dust emission
are negligible.
Free-free emission, however,
can still be appreciable.
We correct the convolved maps
by scaling the WMAP 7-year maximum entropy free-free model
\citep{gold/etal:2011}
to each frequency using spectral index $-2.15$,
convolving the scaled model to 11\fdg6 ~angular resolution,
and subtracting the convolved model from each sky survey.
The resulting maps are dominated by synchrotron emission.

%---------------------------------------------------------
% Figure 2: Sky coverage
%---------------------------------------------------------
\begin{figure}[t]
\includegraphics[width=3.5in]{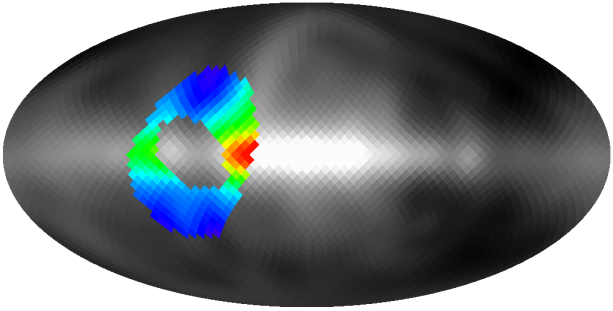}
\caption{Sky coverage for this analysis.
The plot shows the 408 MHz sky survey
convolved to 11\fdg6 ~angular resolution.
Pixels common to all 10 radio surveys
are shown in color.
The sky coverage is limited by the ARCADE 2 observations
but includes
the Galactic plane,
mid-latitude sky,
and portions of the North Galactic Spur (radio Loop I).
\label{sky_coverage_fig} 
}
\end{figure}
%---------------------------------------------------------

\section{Analysis}

The input sky maps define a data set
$T(\hat{n}, \nu)$
sampled at discrete pixel directions $\hat{n}$
and 10 discrete frequencies $\nu$
ranging from 22 MHz to 10 GHz.
We restrict the analysis
to the 8\% of the sky observed at all 10 frequencies.
Figure \ref{sky_coverage_fig} 
shows the resulting sky coverage.
Within the common sky coverage, 
we model synchrotron emission as a modified power law
\begin{equation}
T(\hat{n}, \nu) = A(\hat{n}) 
	\left( \frac{\nu}{\nu_0} 
	\right)^{\beta + C \ln(\nu / \nu_0)}
\label{synch_model}
\end{equation}
with spectral index $\beta$ and curvature $C$
defined with respect to reference frequency 
$\nu_0$ = 310 MHz.
The adopted value for $\nu_0$
minimizes covariance between 
the fitted amplitude $A$ and spectral index $\beta$,
simplifying extrapolation to other frequencies.

For each pixel $\hat{n}$
we define a 10 by 10 data covariance matrix $M$
with diagonal elements
determined by the instrument noise,
calibration and offset uncertainty.
The input maps are not all linearly independent.
Measurements at 22 MHz used the 408 MHz map
to determine the declination dependence of the gain.
We estimate the resulting correlation
of spatial structure in the two maps at 50\%.
The ARCADE 2 maps have independent instrument noise
but share a fraction of the offset uncertainty
related to absolute thermometry uncertainty
and ground glint
\citep{singal/etal:2011}.
All 10 maps share a common model for free-free emission.
We conservatively estimate the uncertainty in the
free-free correction at 30\% of the free-free amplitude;
however,
the results do not change significantly
as the model free-free amplitude is varied by as much as 50\%.
Off-diagonal elements in $M$ include these effects.

%---------------------------------------------------------
% Figure 3: Measured temperatures and best-fit models on Galactic plane
%---------------------------------------------------------
\begin{figure}[b]
\includegraphics[width=2.6in,angle=90]{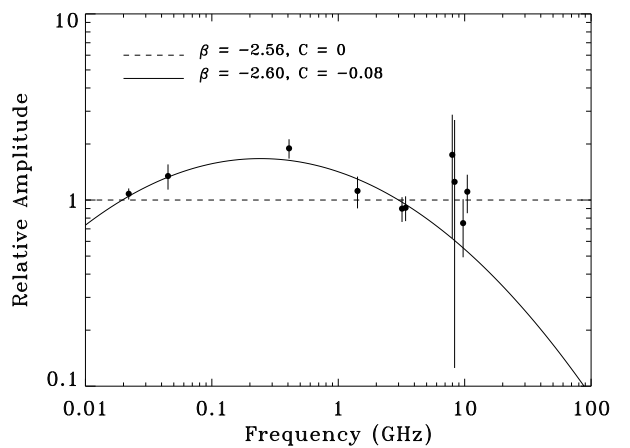}
\caption{Sky temperatures and best-fit model
(solid line)
for a 4$\arcdeg$~diameter patch
on the Galactic plane
centered on the the brightest pixels
in the ARCADE 2 sky coverage.
For clarity, the data are plotted
relative to the best-fit power-law model 
($\beta = -2.56$)
with zero curvature (dashed line).
All fits include the non-trivial covariance
between individual data points.
The data are well described by a model
with spectral index
$\beta = -2.60 \pm 0.04$
and curvature
$C = -0.081 \pm 0.028$.
\label{raw_spectra_fig} 
}
\end{figure}
%---------------------------------------------------------

%---------------------------------------------------------
% Figure 4: Best-fit parameter maps
%---------------------------------------------------------
\begin{figure*}[t]
\includegraphics[width=5.0in,angle=90]{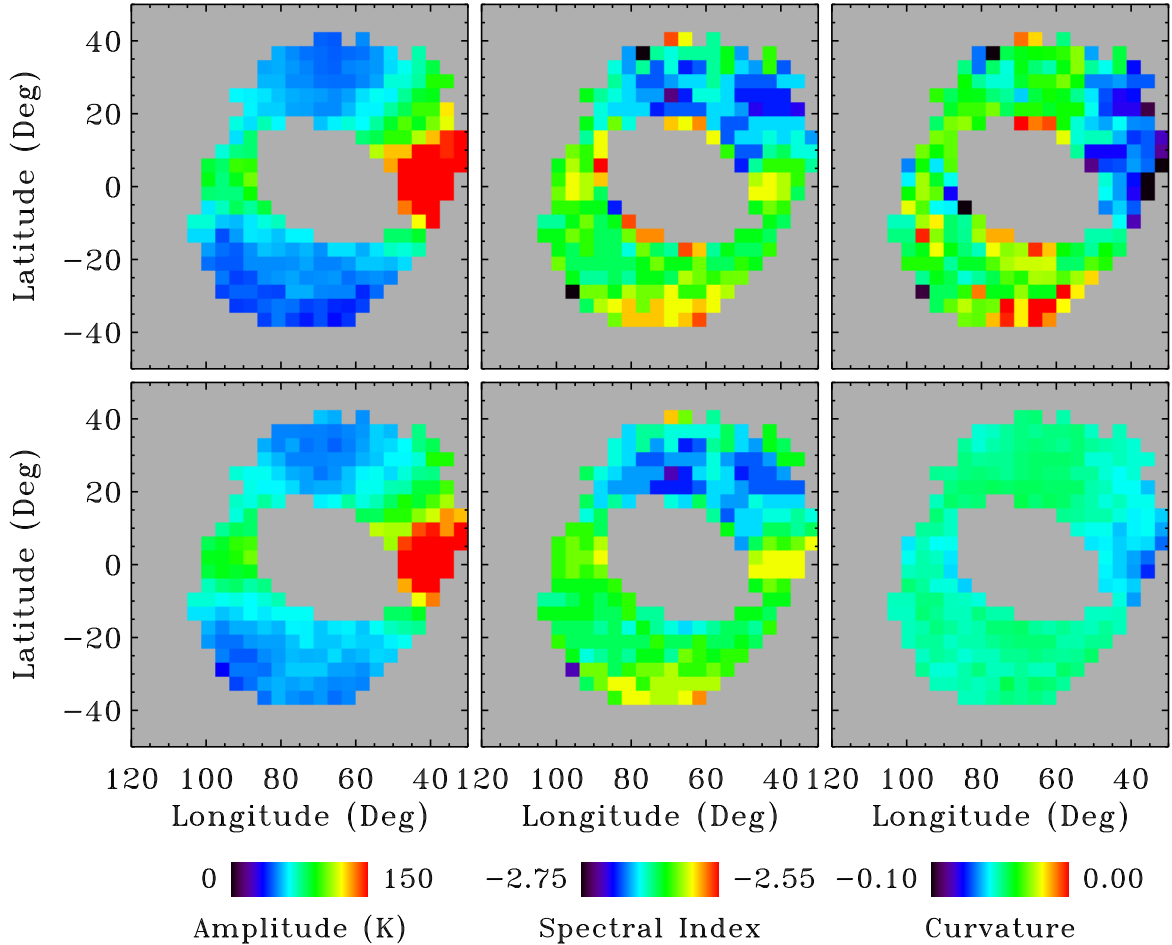}
\caption{Sky maps of best-fit spectral parameters.
Left to right: synchrotron amplitude,
spectral index,
and curvature evaluated at reference frequency 310 MHz.
The top panels show results from the 10-frequency fit,
while the bottom panels
include the spectral constraint at 23 GHz.
\label{param_maps} 
}
\end{figure*}
%---------------------------------------------------------

For each pixel,
a least-squares minimization
determines the best-fit parameters
$A$, $\beta$, and $C$.
Figure \ref{raw_spectra_fig} shows 
the measured temperature and best-fit model
for the brightest Galactic plane region
$(l,b) = (52\arcdeg, 0\arcdeg)$
within the common sky coverage.
The data show evidence for spectral curvature,
with best-fit values
$\beta = -2.60 \pm 0.04$
and
$C = -0.081 \pm 0.028$
evaluated at $\nu_0 = 310$ MHz.
The spectral curvature $C$ for this region is significant at
approximately 3 standard deviations
compared to the baseline model
with $C=0$.

We may extrapolate the spectral models
to compare the results at frequencies 10 GHz and below
to independent determinations of the
spectral index using WMAP data at frequencies 23 GHz and above.
We compute the antenna temperature of the modeled spectra
to derive the effective power-law index
for frequencies near 23 GHz.
Note that this 
is not equivalent to evaluating Eq. \ref{synch_model}
at $\nu = 23$ GHz,
which would yield the scaling from 310 MHz to 23 GHz
but not the power-law index at 23 GHz.
The mean for all 258 pixels in the common sky coverage
is $\beta_{23} = -3.02$
with standard deviation 
$0.22$.

The extrapolated value compares well with
independent determinations of the spectral index above 23 GHz.
\citet{kogut/etal:2004}
analyze WMAP polarization data
to derive synchrotron spectral index $\beta = -3.2 \pm 0.1$
averaged over the full sky.
\citet{dunkley/etal:2009}
use a Bayesian analysis of polarization data
and find the mean spectral index $\beta = -3.03 \pm 0.04$
with pixel-to-pixel standard deviation $0.25$ 
over the high-latitude sky.
\citet{gold/etal:2011}
use template fitting techniques 
to derive spectral index $\beta = -3.13$
between 23 and 33 GHz.

The mean spectral index derived from the 10 low-frequency radio surveys
agrees with the value derived from WMAP data at higher frequencies.
Much of the scatter in the extrapolated spectral indices
results from pixels at high latitude
where the emission is faintest.
The extrapolated index in these pixels
can reach unphysical values.
We reduce the scatter in the fitted spectral parameters
by applying additional constraints using WMAP data at 23 GHz and above.
The simplest such constraint,
adding WMAP temperature data
to the multi-frequency fit,
is problematic.
Not only would the procedure 
need to include additional free parameters
to account for
emission from thermal dust or spinning dust
(both negligible at lower frequencies),
but each of the low-frequency maps
would require a correction to remove 
the monopole contribution
missing from the WMAP data.
Although the WMAP zero level is clearly defined 
by a $\csc|b|$ fit to mid-latitude data,
the astrophysical interpretation of a similar procedure
applied to low-frequency radio surveys is less clear.
The coldest region of the radio sky 
is not at the Galactic poles, 
but at mid-latitudes above the Galactic anti-center.
Subtraction of too large a monopole can leave unphysical
negative residuals.
Limited sky coverage exacerbates this problem.

%---------------------------------------------------------
% Figure 5: Spectral index evaluated at 23 GHz
%---------------------------------------------------------
\begin{figure}[b]
\includegraphics[width=2.7in,angle=90]{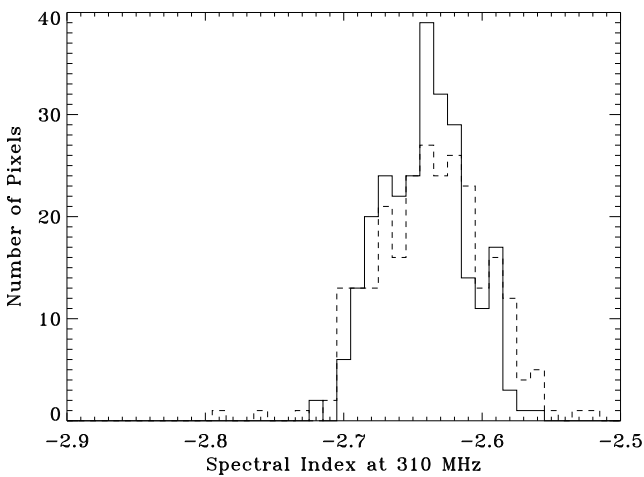}
\caption{Spectral index evaluated at 310 MHz.
The dashed line shows the distribution from the 10-frequency radio data
while the solid line includes the prior at 23 GHz.
\label{beta_hist} 
}
\end{figure}
%---------------------------------------------------------

We avoid these problems 
by using a constraint
based on the spectral index
derived solely from WMAP data.
For each pixel,
we use the radio data
(Table \ref{sky_map_table})
to fit the synchrotron amplitude $A(\hat{n})$
over a 2-dimensional grid in the spectral parameters
$\beta$ and $C$
(Eq. \ref{synch_model}).
At each grid point, we compute the $\chi^2$ value
$ R^T M^{-1} R$
where
$M^{-1}$ is the inverse covariance matrix
and
$R$ is the difference vector between the
measured and modeled temperatures.
We then use the spectral parameters $\beta$ and $C$
to evaluate the power-law index at 23 GHz
and compare the resulting value to a prior.
We use the difference between the extrapolated spectral index
and the prior to augment the $\chi^2$ at each grid point,
\begin{equation}
\chi^2 \rightarrow  \chi^2 + \left( \frac{\beta_{23} - \beta_p}{\sigma_p} \right)^2 ,
\label{delta_chi_eq}
\end{equation}
where
$\beta_{23}$ is the model spectral index
evaluated at 23 GHz,
and 
$\beta_p \pm \sigma_p ~ = ~ -3.1 \pm 0.1$
is the prior at 23 GHz.
The minimum $\chi^2$ over the entire grid
then defines the best-fit model at that pixel.
This allows inclusion of the spectral information
derived from frequencies above 23 GHz
without confusion
from either additional emission components
(thermal or spinning dust) above 23 GHz
or the
missing zero level in the WMAP data.

%---------------------------------------------------------
% Figure 6: Spectral curvature using WMAP prior
%---------------------------------------------------------
\begin{figure}[t]
\includegraphics[width=2.6in,angle=90]{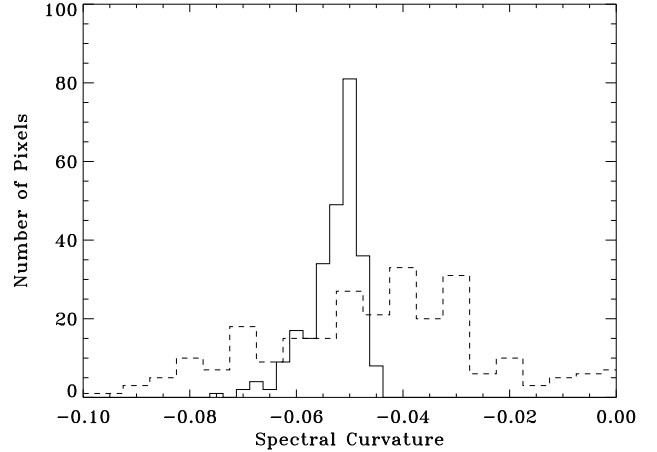}
\caption{Spectral curvature at 310 MHz.
The dashed line shows the distribution from the 10-frequency radio data
while the solid line includes the prior at 23 GHz.
The mean curvature $C = -0.052$ corresponds
to a steepening of the local spectral index
by an amount $\Delta \beta = 0.07$
every octave in frequency.
\label{curv_hist} 
}
\end{figure}
%---------------------------------------------------------

Figure \ref{param_maps} shows the best-fit spectral parameter maps,
while Figures \ref{beta_hist} and
\ref{curv_hist}
show the distribution of the
best-fit values for the spectral index
and curvature.
Including the constraint at 23 GHz,
the best-fit spectral index has 
mean value $-2.64$ 
and standard deviation $0.03$
at reference frequency $\nu_0 = 310$ MHz.
The best-fit curvature has mean value $-0.052$
with standard deviation $0.005$.
The corresponding spectral index at 23 GHz is
$\langle \beta_{23} \rangle = -3.09$
with standard deviation 0.05.
Inclusion of the prior at 23 GHz
does not induce a significant shift in the mean
for the extrapolated spectral index,
but does significantly reduce the pixel-to-pixel scatter.

Comparison of the parameter distributions
with and without the spectral constraint at 23 GHz
demonstrates that the addition of the spectral constraint 
mainly affects the fitted curvature values
(Figs. \ref{beta_hist} and \ref{curv_hist}).
The adopted value for the spectral constraint
comes from independent analysis of WMAP data
and is weighted toward regions of higher synchrotron intensity.
We test whether this creates a bias in the 
fitted curvature values 
by splitting the observed sky coverage
into two subsets
of equal area,
defined by the brightest and faintest 50\%
of the fitted amplitudes at 310 MHz.
Within each subset,
we compare the mean and standard deviation
of the fitted curvature
derived from the 10-frequency fit
or the enhanced fit including the constraint at 23 GHz.
The ``bright'' subset 
shows no shift in the mean spectral curvature
when the 23 GHz constraint is added
(although the scatter is significantly reduced).
The `faint'' subset 
shows both a reduction in scatter
and a modest shift in the mean value,
with curvature parameter steepening from
-0.034 to -0.049 when the 23 GHz constraint is included.
This shift is less than one standard deviation:
given the limited sky coverage,
the available radio data do not yet provide
significant evidence for
spatial variation in the synchrotron curvature.

\section{Discussion}

Radio data show statistically significant steepening 
of the synchrotron spectrum
from 22 MHz to 10 GHz.
The nearly uniform relative uncertainty of the selected data
minimizes dependence of the fitted parameters 
on offset or calibration errors at any one frequency.
We test whether the best-fit parameters
are particularly sensitive to any one input map
by repeating the analysis after dropping one or two maps from the fit.
We select either one radio survey
(22 MHz, 45 MHz, 408 MHz, or 1420 MHz)
or
a pair of ARCADE frequency channels
(3 GHz, 8 GHz, or 10 GHz)
and 
repeat the fit
after deleting the corresponding elements
from the data vector $T$ and covariance matrix $M$.
The resulting shift 
in either the spectral index or curvature parameters
is smaller than the pixel-to-pixel standard deviation
using all 10 frequency channels.
Systematic errors in the offset or temperature calibration
do not appear to dominate the multi-frequency analysis.

The steepening of the synchrotron spectrum
is broadly consistent with models of cosmic ray propagation
in the Galactic magnetic field.
\citet{jaffe/etal:2011} 
combine radio observations
with the {\sc galprop}\footnote{{\tt http://galprop.stanford.edu}}
cosmic ray propagation code
to model synchrotron emission on the Galactic plane.
They find a power-law index
$-2.8 < \beta < -2.74$
from 408 MHz to 2.3 GHz
and
$-2.98 < \beta < -2.91$
from 2.3 GHz to 23 GHz.
The corresponding values 
for the spectral steepening model presented here
are $\beta = -2.76$ from 408 MHz to 2.3 GHz
and $\beta = -2.97$ from 2.3 GHz to 23 GHz,
in good agreement with the cosmic ray model.

We may use the best-fit values in each pixel
to predict the synchrotron spectrum at higher frequencies
where the emission is fainter
and competing sources stronger.
Previous attempts to disentangle 
competing emission
from free-free, synchrotron, thermal dust, 
and spinning dust emission
using WMAP data
have suffered from
degeneracy between
the synchrotron and spinning dust emission,
both of which are falling 
at frequencies above 33 GHz
(see, e.g., the discussion in \citet{gold/etal:2011}).
Extending the synchrotron curvature
observed at lower frequencies
into the millimeter band
reduces confusion between the 
spinning dust and synchrotron spectra
and may facilitate characterization of 
both the spatial distribution and frequency spectrum
of spinning dust emission in the interstellar medium.

%---------------------------------------------------------
% Table 2: Power-law spectral index extrapolation
%---------------------------------------------------------
\begin{table}[t]
\begin{center}
\caption{Local Power-Law Spectral Index}
\label{beta_table}
\begin{tabular}{c c }
\tableline
\tableline
Frequency & Power-Law		\\
(GHz)     & Index		\\
\tableline
  0.022  &	-2.36	\\
  0.045	 &	-2.44	\\
  0.408  &	-2.67	\\
  3.3	 &	-2.89	\\
  23     &      -3.09	\\
  33     &      -3.13	\\
  41     &      -3.15	\\
  61     &      -3.19	\\
  94     &      -3.24	\\ 
\tableline
\end{tabular}
\end{center}
\end{table}
%---------------------------------------------------------

Table \ref{beta_table} shows the 
local power-law index
$T \propto \nu^\beta$
at selected frequency bands.
The modeled spectrum
steepens by $\Delta \beta = 0.07$
every octave in frequency,
from $\beta = -2.67$ at 408 MHz
to $\beta = -3.24$ at 94 GHz.
Note, however, that the spectral steepening observed at
low frequencies can not continue indefinitely.
\emph{Fermi} measurements of the cosmic ray energy spectrum
are consistent with a single power law
from energy 7 GeV to 1 TeV
\citep{abdo/etal:2009,
ackermann/etal:2010,
ackermann/etal:2012}.
If anything, the \emph{Fermi} data suggest 
a modest flattening of the cosmic ray energy spectrum
at higher energies,
which would induce a positive curvature to the synchrotron spectrum
at frequencies above 23 GHz.

%---------------------------------------------------------
% Figure 7: Comparison with Strong et al (2011) model
%---------------------------------------------------------
\begin{figure}[t]
\includegraphics[width=2.7in,angle=90]{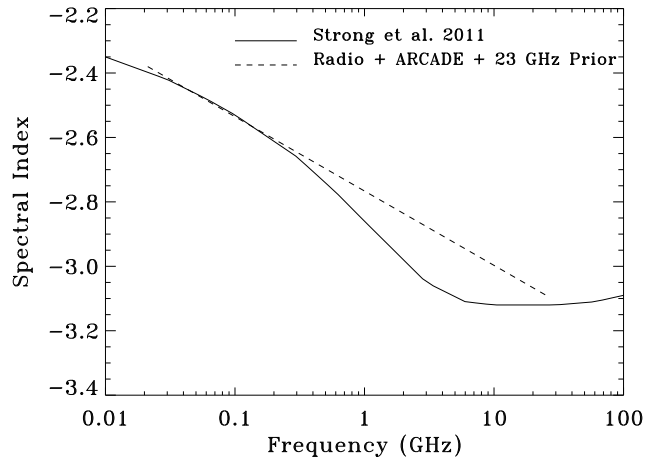}
\caption{Comparison of the 
synchrotron spectral index as a function of frequency
for different models.
The break in the spectral index for the 
\citet{strong/etal:2011} cosmic ray model
at frequencies of a few GHz
is not reproduced by the ARCADE 2 observations
at 3--10 GHz,
which prefer models with nearly constant spectral curvature. 
%
% Add line breaks here to avoid weird preprint pagination
\\
\\
\label{synch_index_comparison} 
}
\end{figure}
%---------------------------------------------------------

\citet{strong/etal:2011} 
combine radio data at
22, 45, 150, 408, and 1420 MHz
with WMAP data at 23 through 94 GHz
and {\it Fermi} Large Area Telescope 
cosmic ray measurements and the {\sc galprop} code
to estimate the magnetic field intensity
and synchrotron spectrum from 22 MHz to 94 GHz.
Figure \ref{synch_index_comparison} compares
the resulting synchrotron spectral index
(diffusion model with injection index 1.3)
to the curvature model from this paper.
The cosmic ray model analyzes a limited latitude range
$10\arcdeg < |b| < 50\arcdeg$,
which we follow in Figure \ref{synch_index_comparison}
by excluding pixels at latitudes $|b| < 10\arcdeg$.
The two methods agree 
for frequencies below 408 MHz
(where they share common radio data)
but differ at higher frequencies.
The cosmic ray model shows an increased synchrotron curvature
from 408 MHz to a few GHz,
followed by a spectral break
to near-constant index $\beta \sim -3.1$
at higher frequencies.
The ARCADE 2 data at 3--10 GHz 
do not reproduce
these features,
but are instead consistent
with constant spectral curvature
from 22 MHz to 10 GHz
(Fig \ref{raw_spectra_fig}).

Both models agree at frequencies near 23 GHz.
The spectral index at 23 GHz
derived from the 10-frequency radio data
{\it without} the 23 GHz prior
is consistent with
independent measurements of the index above 23 GHz
and with the full radio fit including the 23 GHz prior.
The radio data, taken alone,
do not support a break in the synchrotron spectrum at GHz frequencies.
Comparison of the radio fit
to the cosmic ray model
suggests that any spectral break in the synchrotron spectrum
must occur at frequencies above 23 GHz.
Direct confirmation of the synchrotron spectrum
above 23 GHz remains a challenge.

% Add somce space here to avoid weird pagination
\vspace{5mm}
\section{Conclusions}

Radio data are consistent with a synchrotron spectrum
that steepens with frequency from 22 MHz to 10 GHz.
Direct comparison of low-frequency radio surveys
with the WMAP data at 23 to 94 GHz
is complicated 
both by the presence of additional emission components
at higher frequencies
and by the subtraction of a substantial monopole component
of sky emission
by the differential WMAP instrument.
The synchrotron spectral index at 23 GHz,
derived using only lower-frequency radio surveys,
is consistent with the value derived independently
using only data at higher frequencies.
We extend the radio data
by comparing the extrapolated index at 23 GHz
to a prior based on higher-frequency data.
The combined data have mean spectral index
$\beta = -2.64 \pm 0.03$ 
and curvature
$C = -0.052 \pm 0.005$
at reference frequency 0.31 GHz.
The measured spectrum steepens
by an amount
$\Delta \beta = 0.07$ every octave in frequency.
Comparison of the radio data to models 
including the cosmic ray energy spectrum
suggests that any break in the synchrotron spectrum
must occur at frequencies above 23 GHz.

\acknowledgements

This research is based upon work supported by
the National Aeronautics and Space Administration
through the Science Mission Directorate
under the Astronomy and Physics Research and Analysis 
suborbital program.

% Add page break here to clear weird page formatting 
% \clearpage

% --------------------- References  ---------------------

% Exit, stage left
\end{document}